# Low-loss multilevel operation using lossy PCM-integrated silicon photonics

Rui Chen[a,*], Virat Tara[a], Jayita Dutta[a], Zhuoran Fang[a], Jiajiu Zheng[a], Arka Majumdar[a,b,*]

[a]Department of Electrical and Computer Engineering, University of Washington, Seattle, WA 98195, USA
[b]Department of Physics, University of Washington, Seattle, WA 98195, USA

**Abstract**. Chalcogenide phase-change materials (PCMs) offer new paradigms for programmable photonic integrated circuits (PICs) thanks to their zero static energy and significant refractive index contrast. However, prototypical PCMs, such as GeSbTe (GST), are lossy in their crystalline phase, albeit transparent in the amorphous state. Moreover, electrically switching PCMs to intermediate states is a stochastic process, limiting programming accuracy. As a result, achieving both low-loss and deterministic multi-level operation with GST remains challenging. Although low-loss PCMs, such as $Sb_2S_3$ and $Sb_2Se_3$, have been discovered in recent years, they are much less technologically mature. In this work, we propose a design with multiple GST segments to overcome the challenge of deterministic multilevel operation. GST segments are individually controlled by interleaved silicon PIN diode heaters in a binary but reliable fashion, and multiple levels are encoded in their phase sequence. A 1 × 1 programmable unit with two unequal GST segments is experimentally demonstrated, showcasing four distinct operation levels and negligible thermal crosstalk with only one pair of metal contacts. We then extend the design to 1 × 2 and 2 × 2 programmable units. For the 2 × 2 programmable unit design, we propose a phase-detuned three-waveguide directional coupler structure to mitigate the absorption and radiation loss, showing < -1.2 dB loss and three splitting ratios. Our work provides a new path toward low-loss and multi-level optical switches using lossy PCMs.

*Rui Chen, E-mail: charey@uw.edu; Arka Majumdar, E-mail: arka@uw.edu

## 1 Introduction

Programmable photonic integrated circuits (PICs), consisting of meshes of tunable beam splitters and/or optical phase shifters,[1] are extending their applications from widely used optical communication to optical computing,[2] microwave photonics[3] and quantum information processing.[4,5] To further scale up the programmable PICs, it is necessary to achieve low loss, low crosstalk, compact, low power consumption, multi-level programmable units. However, most programmable PICs utilize weak and volatile tuning mechanisms such as thermo-optic,[6,7] electro-optics[8] or free-carrier effects,[9] providing very small and volatile refractive index change ($\Delta n$ < 0.01). As a result, the devices usually have a large footprint (> 100 μm[10]) and require a constant power supply (~10 mW), incurring limited density and energy efficiency.

Nonvolatile chalcogenide phase-change materials (PCMs)[11–14] are a very promising platform. These materials have two stable micro-structural phases in the ambient environment, *i.e.*, the thermodynamically favorable crystalline- (c-) phase and the meta-stable amorphous- (a-) phase. Due to the change in atomic arrangement, a- and c- phase differ significantly in their refractive index ($\Delta n \sim 1$). Earlier experiments have shown that the phases can be controlled both optically using pulsed lasers[15–18] and electrically via on-chip microheaters.[19–23] It is worth noting that since both phases are stable, zero power is needed to maintain the state once the PCM phase is changed. In addition, this memory effect can reduce the complexity of control circuits as electronic memory blocks and time-multiplexing are avoided. The large $\Delta n$ and nonvolatility together offer compact devices (~10 μm[23,24]) with zero static energy consumption, hence a unique opportunity to achieve ultra-low-power programmable PICs.

Most technologically mature PCMs, such as $Ge_2Sb_2Te_5$ (GST) and GeTe, are very lossy in the c-phase particularly in the near infrared regime (~1550nm), although they are transparent in the a-phase. In spite of the emergence of wide bandgap PCMs, such as $Sb_2S_3$ and $Sb_2Se_3$,[21–23,25–28] they are much less understood and therefore are less likely to be integrated in large-scale foundry processes. In this work, we consider the prototypical PCM GST, which has been widely used for electronic memory applications[29,30] and holds promise to integrate into the photonic foundries. As explained earlier, one major limitation of GST at 1,550 nm is the absorption loss of c-GST, which prohibits it from phase-only modulation as needed for large-scale switching fabrics[31] or interferometry-based optical signal processors.[2] To mitigate the loss, a phase-matched three-waveguide directional coupler (PM-TDC) design was reported in the previous studies[20,24,31]. In addition to the loss, deterministic multi-level operation also remains challenging. Although pulse width or amplitude modulation[22,23] could produce mixed amorphous-crystalline phase, hence multiple operation levels, such process is inherently stochastic.[32] Furthermore, these schemes cannot be used in the PM-TDC design, which causes a large loss due to either material absorption or radiation loss at the end of the transition waveguide.

In this work, we propose and numerically demonstrate low-loss optical programmable units based on a lossy PCM GST, including 1 × 1, 1 × 2 and 2 × 2 programmable units. Our devices support deterministic multi-level operation with individually controlled GST segments, which are controlled by a series of interleaved p++-doped-intrinsic-n++-doped (PIN) silicon microheaters. Each microheater controls the corresponding segment in a binary way, and different states can be achieved by programming the phases of the GST segments. We emphasize that this scheme is inherently deterministic compared to engineering the electrical pulse amplitude[22], duration or number of pulses[21], as it solely relies on the binary phase transition, which is more deterministic than achieving intermediate states. We show that 1 × 1, 1 × 2 programmable units can achieve multiple operation levels using segmented PCM. We reported two methods to achieve $2^N$ and ($N$+1) states with $N$ segments. A preliminary experiment on 1 × 1 waveguide programmable units with the segmented GST shows distinct multiple operation levels without significant thermal crosstalk. The rectified property of PIN diode heaters was exploited in this programmable unit to reduce the number of metal contacts to only one pair to give four operational levels. Numerical simulation shows that this method can be easily extended to 1 × 2 programmable units. However, 2 × 2 programmable units based on the same method show very high loss in the intermediate levels. We model the PM-TDCs both analytically and numerically to unveil that it is the fundamental wave dynamics in coupled-waveguide systems that causes a combination of material absorption loss (Type I) and radiation loss at the end of the transition waveguide (Type II). In a PM-TDC design, either one of two types of loss presents for intermediate states, incurring high optical loss.

A modified structure, termed here as phase-detuned three-waveguide directional coupler (PD-TDC), is proposed to mitigate both types of loss, where the center transition waveguide (WG) is deliberately phase detuned from the other two WGs for a-GST. By programming the GST patches, we numerically show a three-level 2 × 2 programmable unit with low insertion loss (~ -1.2 dB). Our work provides a new design methodology for programmable photonic devices using lossy PCMs or any other modulation mechanism.

## 2 Deterministic multi-level scheme using individually controlled GST segments

Figure 1 shows our scheme to achieve deterministic multi-level operation using binary phase transition of PCM. Instead of engineering the electrical pulse conditions[21,22], we propose to individually control multiple segments of GST thin films on silicon WGs to either fully a- or c-phase using multiple PIN diode heaters with opposite polarity[20,21,33]. By programming the state of each GST segment, multiple operation levels can be obtained.

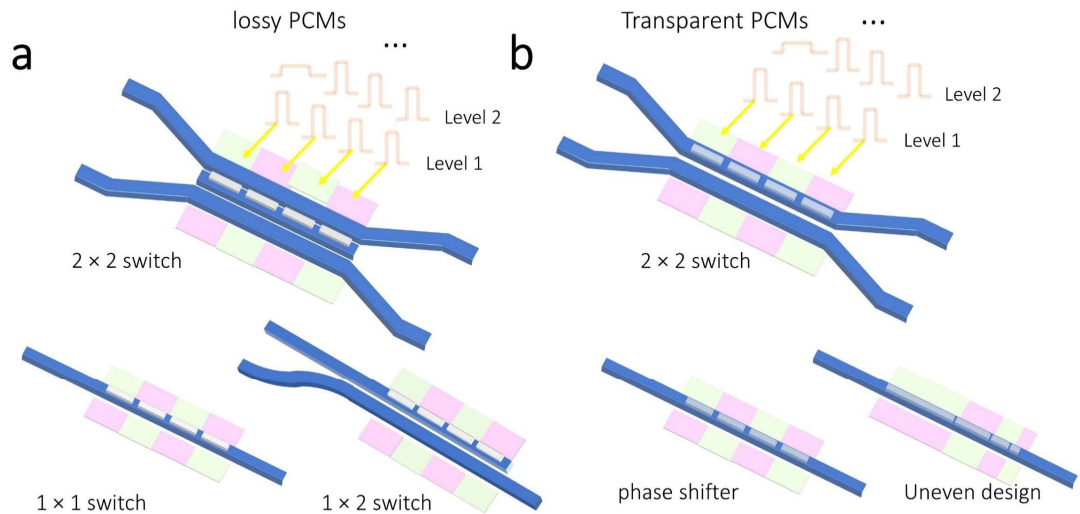

**Fig. 1** Our proposed deterministic multi-level programmable units, with multiple PIN diode heaters with reversed polarity to control the phase of each PCM segment in a binary manner: (a) programmable units using lossy PCMs,

and (b) programmable units using transparent PCMs. An example of uneven segment design is shown at the bottom right. (Colors: pink: p++ doped silicon; green: n++ doped silicon; blue: intrinsic silicon; white: lossy PCMs; gray: transparent PCMs.)

## 3 Experimental demonstration of multi-level 1 × 1 programmable units with two unequal segments

A multi-level 1 × 1 WG programmable unit can be implemented for either amplitude or phase modulation, depending on whether the PCM is lossy in the crystalline phase. At telecommunication wavelength of 1,550 nm, GST,[19] GSSe[34] or GSST[35] can be used for amplitude modulation, and $Sb_2Se_3$[22,23] or $Sb_2S_3$[21,26,36] can realize phase-only modulation. In the design in Fig. 1a, each GST segment represents 1 bit, and a device with $N$ segments can present $2^N$ bits at most. This can be achieved by designing the lengths of GST segment $i$ twice that of segment $i$-$1$, see the bottom right sketch in Fig. 1. We note that in general unequal segments can generate $2^N$ levels, but the difference between levels is not the same unless the lengths satisfy the double-length condition.

The rectification behavior of PIN diode heaters can be leveraged to reduce the number of metal wires. By using a series of alternating PIN heaters (Fig. 1a), two PIN diode heaters can share the same metal pad, because reversely biased current is small below the breakdown voltage and not enough to trigger any phase transition. The number of metal pads can be reduced by half, hence simplified electrical wire routing. This is beyond the capability of traditional resistive heaters such as the tungsten heater in Ref. [34].

One potential limitation is the thermal crosstalk between adjacent GST segments. We experimentally show that this thermal crosstalk does not trigger unintentional phase transition by fabricating and characterizing a 1 × 1 programmable unit. The fabrication process is similar to our previous works[20,21,28]. We designed the programmable unit with two unequal GST segments (10-nm-thick, 1.67-μm- and 3.34-μm-long) and demonstrated four distinct levels. Figure 2 shows the device layout, fabricated devices, and characterization results. The lengths of the PCM patches are engineered (Fig. 2b) to provide a maximum number of $2^N$ states ($N = 2$, therefore 4 levels). The fabricated device is shown in Fig. 2c. Time trace sequences in Fig. 2d were obtained by recording the transmission at 1,550 nm while alternatively applying amorphization (4 V, 100 ns) and crystallization pulses (1.4 V, 50 μs) on Segment 2. The on-off extinction ratio (ER) is 0.6 dB, working as a high-precision tuning knob. Low-precision tuning (T0 to T1 and T1 to T2, on-off ER of ~3 dB) was realized by sending in pulses in a reversed bias to control the longer GST segment 1, with -4.5 V for amorphization and -1.5 V for crystallization. We note the polarity is reversed compared to the longer segment. This result shows that the short and long segments are separately tuned without noticeable thermal crosstalk.

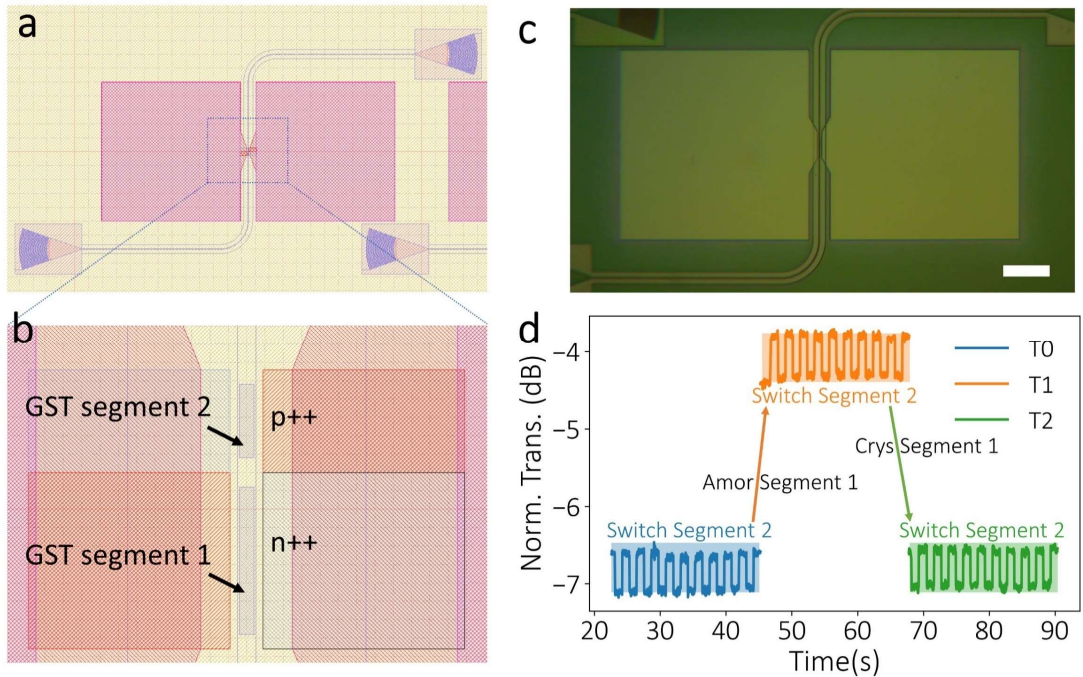

**Fig. 2** Deterministic four-level 1 × 1 programmable unit using two PIN heaters with reversed polarity: (a) Schematic of the device. (b) Zoomed-in schematic. Scalebar: 25 μm. (c) Optical microscope image of the fabricated device. (d) Time-trace measurement, showing four distinct and reversible transmission levels. Measured results for time trace

T0, T1 and T2 are in blue, orange and green, respectively, where fine tuning is achieved in each time trace by switching Segment 2 and coarse tuning is shown between two adjacent time traces by switching Segment 1.

## 4 Optical design for deterministic multi-level 1 × 2 programmable unit

The experimental result in Sec. 3 validates our deterministic multi-level scheme in 1 × 1 programmable unit. In this section, we extend the idea to 1 × 2 and 2 × 2 programmable units, which require phase-only modulation and are building blocks for large-scale programmable PICs. To achieve a low-loss operation, a three-waveguide directional coupler consisting of GST-loaded silicon (hybrid) WG and the bare silicon WG was previously reported.[20,24,31] When GST is in the amorphous phase, the hybrid WG is designed to phase match with the bare silicon WGs. Therefore, the input light couples to the cross port. When the GST changes to the crystalline state, due to the large effective index difference, the WGs are phase mismatched. As a result, the light barely couples and stay in the bar port, thus bypassing the high loss of c-GST. We emphasize that achieving multi-level operation using this idea requires extra design considerations, since the absorption loss of c-GST limits the available phase sequence of the GST segments.

Fig. 3 shows the simulation result for a 1 × 2 optical programmable unit, for which the design is the same as Ref. [20]. One change is that two individually controlled GST segments, instead of one, are used to achieve multi-level operation. The achievable low-loss states are, however, not $2^N$ as in the 1 × 1 programmable unit case, but $N + 1$, since any a-GST segment before c-GST segment results in light coupling to the GST-loaded waveguide, which then gets completely absorbed by the highly lossy c-GST segments. Our simulation results agree with this intuition. We denote the phase sequence from left to right as {phase of segment 1, phase of segment 2} and show the electric field distribution and the transmission spectrum for all four possible phase sequences. Figure 3b shows that Sequence {a, c} exhibits a high insertion loss of ~ -3 dB due to high absorption of the c-GST segment 2. This is verified by the electric field distribution image, where the light first couples into the cross port along the a-GST segment 1, then gets completely absorbed by c-GST segment 2. Since the length of the first segment is half of the coupling length, exactly half of the light is coupled and absorbed, agreeing well with the simulation.

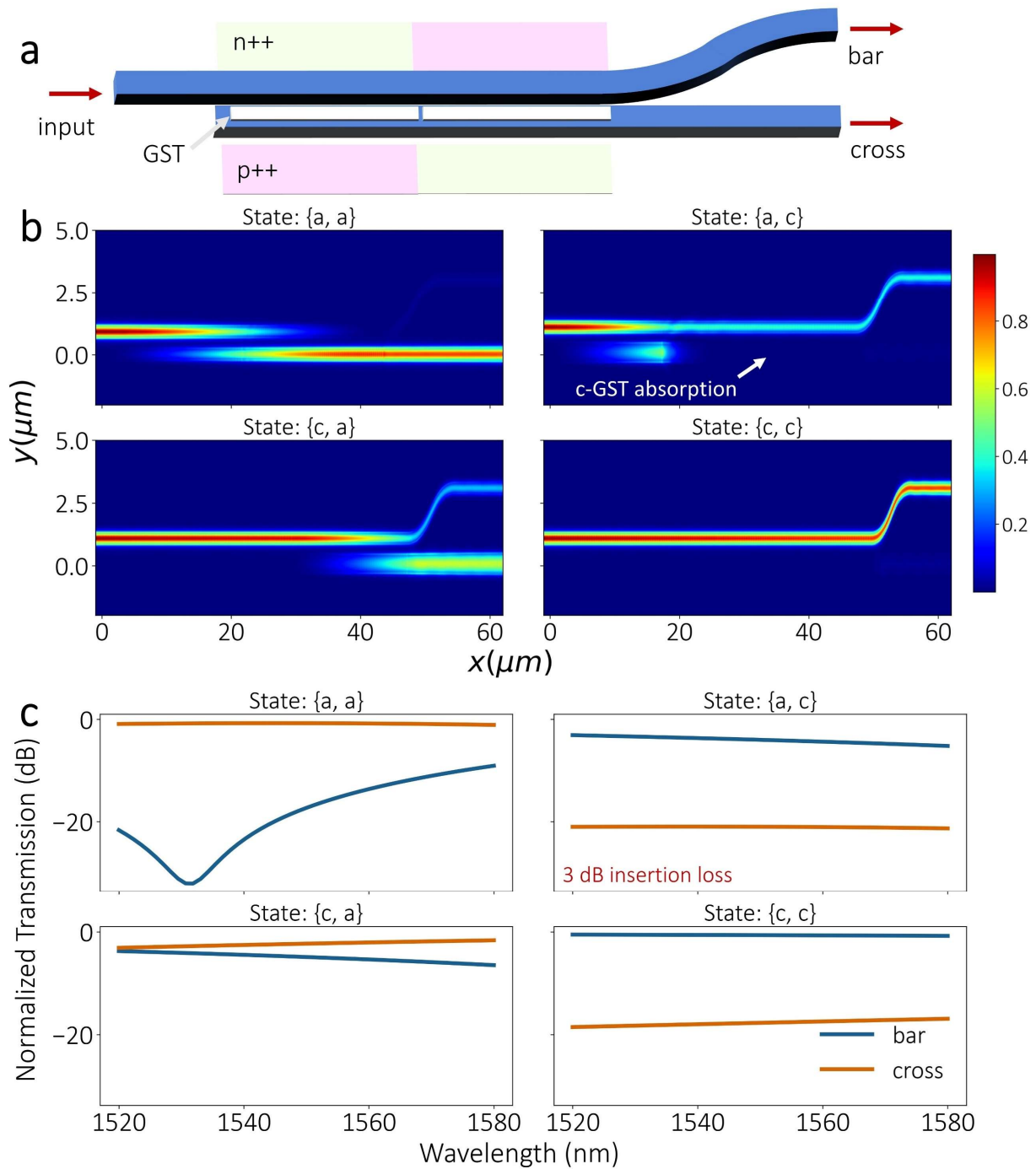

**Fig. 3** FDTD simulation for a deterministic three-level 1 × 2 programmable unit using GST: (a) Schematic. (b) Intensity distribution and (c) the corresponding transmission spectra for different GST phase combinations. GST sequence {a, c} introduces an insertion loss of ~ -3 dB due to c-GST absorption loss, pointed by a white arrow in the field distribution map.

## 5  Optical design for deterministic multi-level 2 × 2 programmable unit

Previous GST-based low-loss 2 × 2 programmable unit leveraged a PM-TDC structure to circumvent the loss of c-GST.[20,24,31] In these works, an extra transition WG is placed between two silicon waveguides with 20-nm-thick GST deposited on top for active switching. When the GST is in the a-phase, the transition WG is designed to phase match with other two WGs. Therefore, the input light can be coupled into the transition WG, and then to the cross port. On the contrary, once switching the GST to the c-phase, a large effective index difference is introduced, and the light can barely couple to the high loss transition WG. This bypasses the high loss of c-GST. Unfortunately, a straightforward extension of this PM-TDC method to multi-level operation does not work as we will show in the following analysis.

*5.1 Challenges of multiple segments switching using the phase-matched TDC approach*

Figure 4 shows a simulation using a similar design to Ref. [20] but with two GST segments. It shows that both {a, c} and {c, a} phase sequences have a high loss of ~ -3 dB. The former can be explained based on the same concept as in the 1 × 2 programmable unit case: no c-GST should be placed after an a-GST segment when finite light is still in the transition WG, defined as the Type I absorption loss. Besides, the transition WG, which is open at the end, adds another constraint for low-loss operation: no light can remain in the transition WG at the end of this transition waveguide. Otherwise, that portion of light will radiate and get lost, defined as Type II radiation loss.

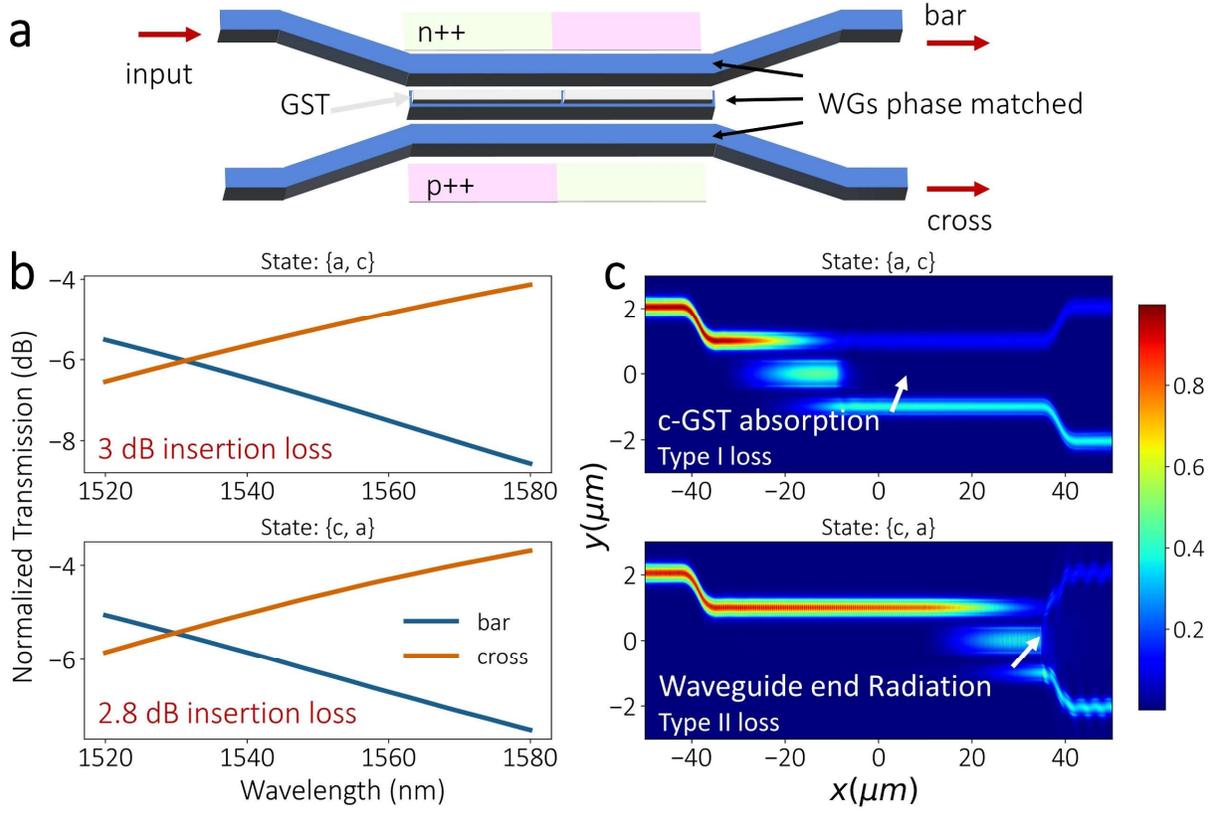

**Fig. 4** Phase-matched three-WG 2 × 2 directional couplers have high loss for intermediate levels: (a) Schematic. (b) the transmission spectra for different GST phase sequences and (c) the corresponding intensity distribution. Both sequences {a, c} and {c, a} introduce an insertion loss of ~ -3 dB. The former is due to c-GST absorption loss (Type I loss), while the latter is due to the radiation loss at the end of the transition WG (Type II loss), both indicated by white arrows.

*5.2 Analytical and numerical study of the three-waveguide directional coupler systems*

To tackle this loss issue, it is crucial to understand the intensity dynamics inside of each WG in this coupled-waveguide system. We describe the coupled three waveguide system using the following coupled partial differential equation[37]

$$\begin{cases} \frac{da_0}{dz} = i\kappa^*(a_1 + a_2) \\ \frac{da_1}{dz} = i\kappa a_0 - i\delta a_1 \\ \frac{da_2}{dz} = i\kappa a_0 - i\delta a_2 \end{cases}, \tag{1}$$

where $z$ is the distance in light propagation direction, $a_{i,i=0,1,2}$ is the amplitude in three WGs in Fig. 5a, $\kappa = \Delta n_{eff,c} \cdot \frac{2\pi}{\lambda_0}$ and $\delta = \Delta n_{eff,i} \cdot \frac{2\pi}{\lambda_0}$ are the coupling coefficient and detuning between the center and other WGs, $\Delta n_{eff,c}$ is the effective index difference between the super-modes formed by coupled WG0 and WG1 (or WG2), $\Delta n_{eff,i}$ is the effective index difference between the modes in WG0 and WG1 (or WG2) assuming they are isolated, and $\lambda_0$ is the vacuum wavelength of light. We note that only the nearest coupling is considered in this model, which is usually a valid assumption in a 2D photonic WG array. We also assume that WG1 and WG2 have the same geometry to simplify the model and neglect the loss term for simplicity. We note that this coupled equation is analytically solvable, see Appendix I. To quickly capture the physics, we show the numerical analysis below.

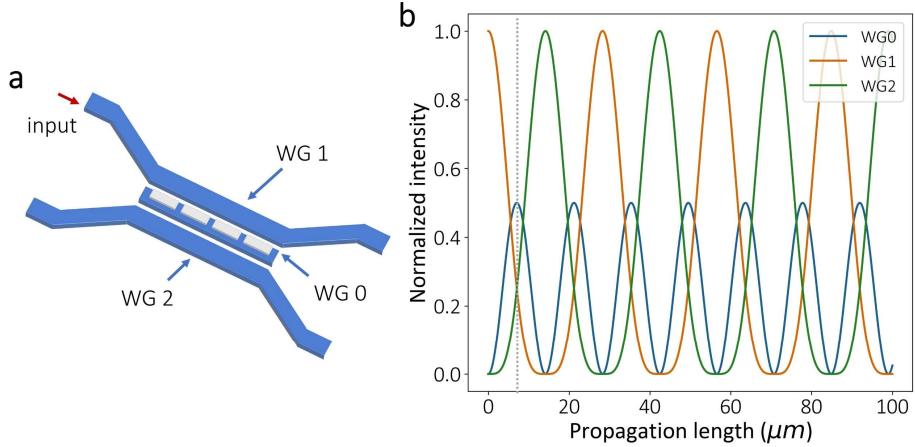

**Fig. 5** Analytical model reveal that phase-matched three-WG couplers can never have low-loss intermediate levels: (a) Schematic for a generic GST three-waveguide coupler. The GST patches are denoted in white. In all the discussions, we assume the WG1 the input port. (b) Numerically calculated light intensity in three WGs. Blue, orange and green colors denote WG0, WG1, and WG2, respectively. In the case of two GST segments, phase sequence {c, a} would lead to a propagation distance of $L_c/2$, indicated by the gray dotted line. It is clear that finite energy still remains in WG0, causing Type II radiative loss.

First, to analyze the case where WG0 is phase matched with WGs 1 and 2, we set $\delta$ to zero and solve this equation numerically. We found that the middle WG intensity is zero only when the light is completely in the cross or bar port (Fig. 5b). Therefore, Type I loss occurs in the {a, c} state, while Type II loss occurs in the {c, a} state. Therefore, we summarize the requirements for low-loss multilevel operation in a three-waveguide directional coupler as (1) no light can be in the middle waveguide at a c-GST segments to avoid Type-I loss, and (2) no light can be at the end of the middle waveguide to avoid Type-II loss, and (3) there must exist some intermediate splitting ratios (except from complete bar or cross) when the intensity of light in WG0 is zero. Obviously, the last requirement can never be fulfilled in a phase-matched TDC.

*5.3 New degree of freedom: phase-detuned three-waveguide directional couplers*

To solve this issue, a new degree of freedom must be introduced. We deliberately designed WG0 to have a phase detuning $\delta$ with WG1 and WG2, *i.e.*, a phase-detuned TDC scheme. The phase detuning between WGs can be quantitatively represented by the detuning rate $\delta$, where $\beta = 2\pi/\lambda$ is the vacuum wavevector and $\Delta n_{eff}$ is the effective index difference between WGs. To intuitively understand the wave dynamics in this system, we performed a numerical simulation as shown in Fig. 6, where different plots show the intensity in the three WGs versus propagation length under different detuning rates $\delta$. It can be clearly seen that different splitting ratios between WG1 and WG2 can be obtained when zero light is in WG0. We note that Eqns. 1 is analytically solvable, and the solution is presented in Appendix I.

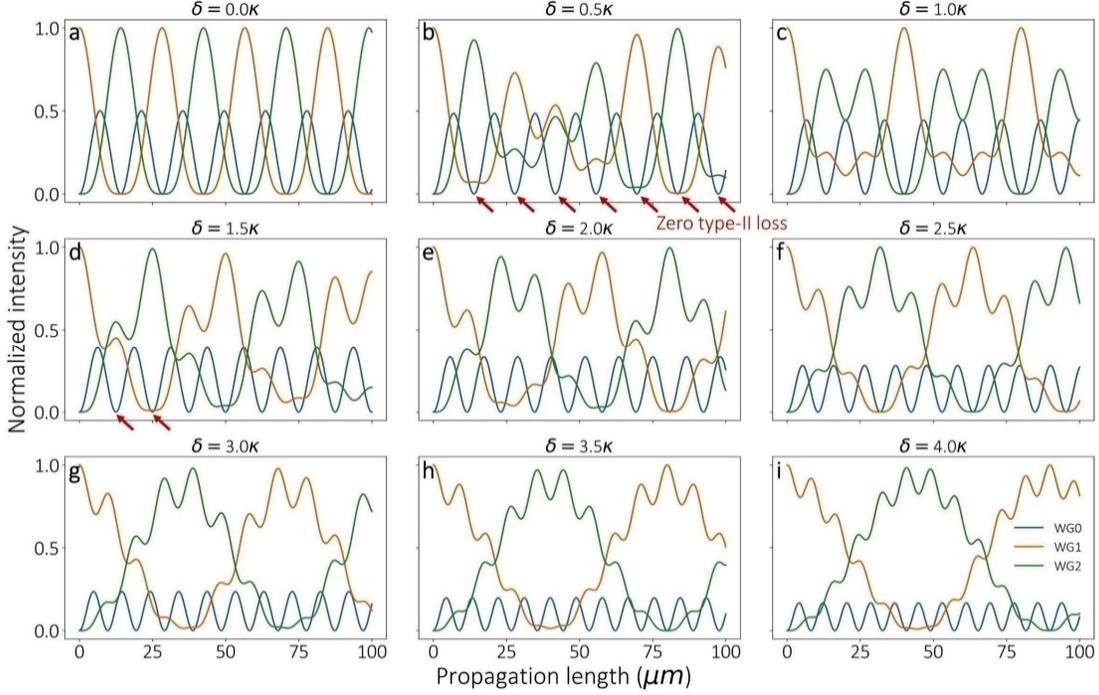

**Fig. 6** Detuning the middle waveguide provides rich optical intensity dynamics, enabling low-loss intermediate levels: (**a**)-(**i**) Intensity dynamics in the three waveguides as the detuning $\delta$ varies from 0 to $4\kappa$. As the detuning gets larger, the intensity in WG0 becomes smaller and that in the other two is smoother. As $\delta$ increases from zero, multiple zeros in blue curve emerge (such as pointed by red arrows in **b**), overcoming Type II loss. Note that this simulation assumes a coupling length $L_c$ of 10 $\mu m$ between WG0 and WG1 (or WG2) when $\delta = 0$.

A tradeoff must be considered when choosing detuning rate $\delta$. A large detuning $\delta$ generally creates more zeros in the middle waveguide within the same coupling region length, hence more achievable intermediate levels. However, a large $\delta$ also leads to a longer device (for complete power transfer to the cross port) and hence larger overall loss when GST is in c-phase. To compensate for this, a thicker GST can be used to provide a larger $\Delta n$. However, the loss due to a-GST also increases. Furthermore, this also makes the reversible electrical switching more difficult.

## 5.4 A design example of nonvolatile multi-level 2 × 2 programmable units based on GST

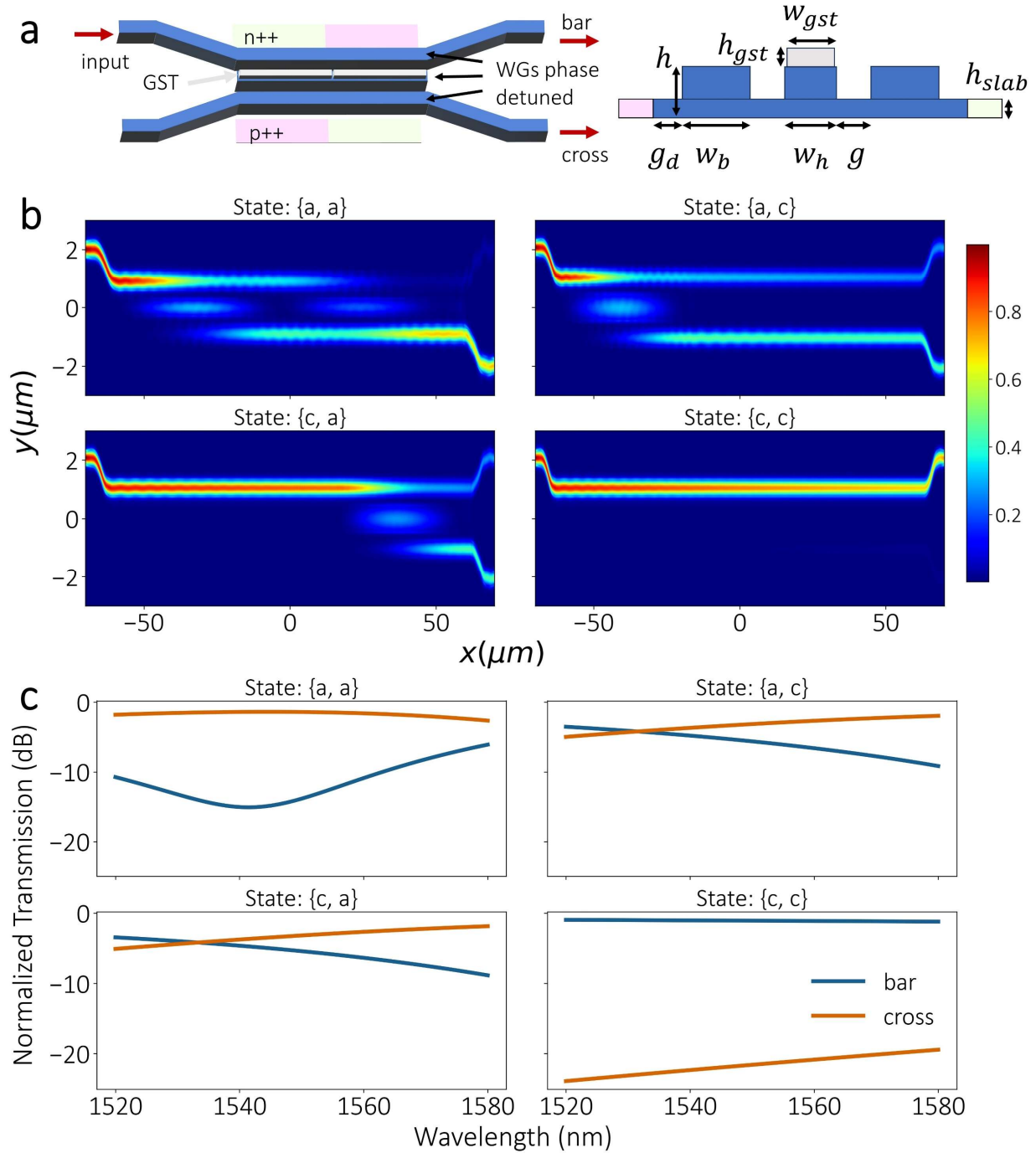

**Fig. 7** Optimized deterministic three-level 2 × 2 programmable units with phase-detuned TDC design ($\delta = 1.5\kappa$): (a) Schematic (left) and the cross-sectional view (right). Designed geometry parameters are listed in Table 1. (b)

Simulated intensity distribution and (c) the corresponding transmission spectra for different GST phase sequences. The insertion loss is reduced to ~-1.2 dB compared to ~ -3 dB in traditional phase-matched TDC designs.

Based on the above discussion, we choose $\delta = 1.5\kappa$ and design a multi-level 2 × 2 programmable unit with ellipsometry measured GST refractive index, (see Appendix II for detailed design procedures). Figure 7 shows the FDTD simulation results with a low insertion loss (~ -1.2 dB) and three achievable splitting ratios of 100:0, 50:50 and 0:100. The designed parameters are summarized in Table 1 below.

**Table 1** Designed parameters for the multi-level 2 × 2 GST programmable unit with $\delta = 1.5\kappa$. Unit: nm except for $l_c$.

| $h$ | $h_{slab}$ | $g_d$ | $w_b$ | $w_h$ | $g$ | $h_{gst}$ | $w_{gst}$ | $l_c$ (μm) |
|---|---|---|---|---|---|---|---|---|
| 220 | 100 | 200 | 630 | 500 | 315 | 30 | 450 | 115 |

Figure 8 shows the insertion loss and extinction ratio of the optimized 2 × 2 programmable unit. The insertion loss is low (~-1.2 dB) across a broad wavelength (60 nm) for all three levels (four configurations). We emphasize that this insertion loss is mainly due to the relatively thick GST film used in the design, which introduces absorption loss in its amorphous state. By replacing GST with lower loss PCMs such as GSST, we can potentially further reduce the loss. However, the index contrast $\Delta n$ of GSST is much smaller than GST, which may result in a thicker PCM and longer devices. The extinction ratios of the complete cross and bar states remain high across 60 nm, but the intermediate states are more wavelength-sensitive with a 3 dB variation bandwidth of ~ 22.6 nm.

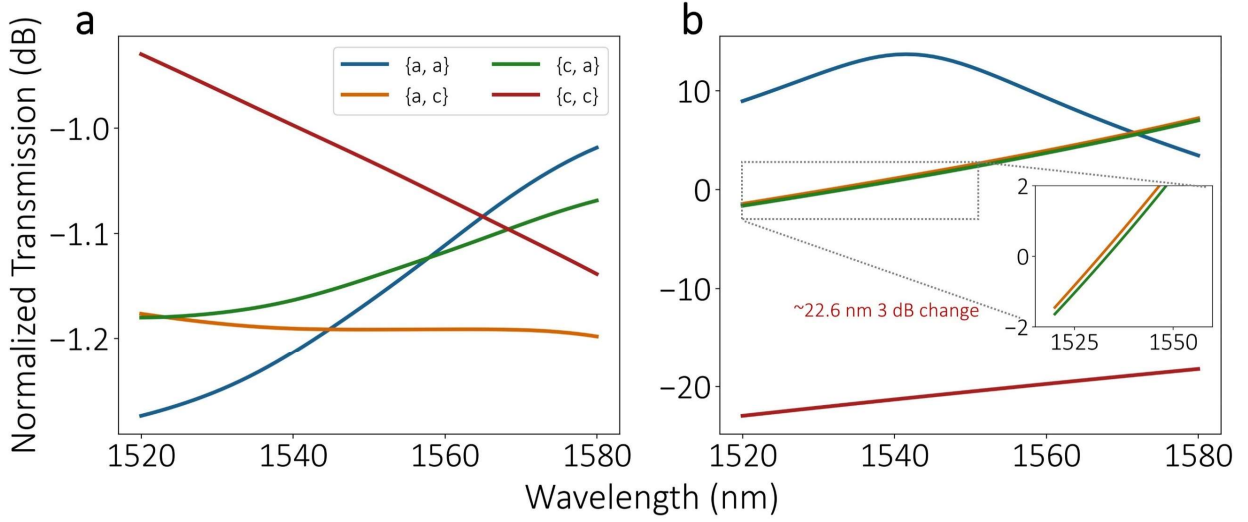

**Fig. 8** Insertion loss and extinction ratio of the designed three-level 2 × 2 TDC programmable unit ($\delta = 1.5\kappa$): The device shows a low loss (< -1.2 dB) for all four GST sequences and a bandwidth of ~22.6 nm with less than 3 dB extinction ratio variation.

## 7 Discussion

We emphasize that using transparent PCMs, such as $Sb_2Se_3$ and $Sb_2S_3$, can easily extend the multi-level design to a 2 × 2 programmable unit.[21,38] The design principle is similar, and the switching of output ports relies on the effective index change of the hybrid waveguide when the PCM is switched. Importantly, since transparent PCMs introduce negligible excess loss, $2^N$ distinguished splitting ratios can be easily achieved by varying PCM segment lengths. Our recent demonstration shows the promise of using a set-and-verify method to achieve accurate operation levels[21], but it also requires significant time and power overhead for fine-tuning. These transparent PCMs, however, are more challenging to be integrated into photonic foundries for massive production.

In summary, we have numerically demonstrated GST-based silicon photonic programmable units with low insertion loss and deterministic multi-level. The multi-level operation is achieved via a novel design where interleaved PIN silicon heaters separately control each GST segment. By programming each GST segment's phase, deterministic multilevel operations are obtained. To demonstrate that the thermal crosstalk does not introduce unwanted PCM phase transition, we fabricated and characterized a two-GST-segment, four-level 1 × 1 programmable unit. The experimental result shows four distinct transmission levels, and each level is accessible, reliable, and reversible. Based on this idea, we also designed 1 × 2 and 2 × 2 programmable unit with low insertion loss. A phase-detuned TDC geometry was proposed to achieve low-loss 2 × 2 programmable unit and we numerically demonstrated a design with a low insertion loss (~-1.2 dB) and three operation levels. This novel design idea provides a way to achieve low-loss photonic switches with multiple operation levels using lossy tuning media.

**Appendix A: Low-loss multi-level 2 × 2 GST programmable unit design procedure**

*A.1 Analytical solution to Eqs. (1)*

Equation (1) is analytically solvable. The light intensity inside of each WG can be expressed as

$$\begin{cases} I_0(z) = \frac{1}{\left(\frac{\delta}{2\kappa}\right)^2 + 2}\left[1 - \frac{\cos(\sqrt{\Delta}z)}{2}\right] \\ I_1(z) = \frac{1}{4}\left[1 + \frac{4\kappa^2 + \delta^2}{\Delta} + \frac{\sqrt{\Delta}-\delta}{\sqrt{\Delta}}\cos(\beta_{12}z) - \frac{\sqrt{\Delta}+\delta}{\sqrt{\Delta}}\cos(\beta_{13}z) - \frac{\Delta-\delta^2}{\Delta}\cos(\beta_{23}z)\right], \\ I_2(z) = \frac{1}{4}\left[1 + \frac{4\kappa^2 + \delta^2}{\Delta} - \frac{\sqrt{\Delta}-\delta}{\sqrt{\Delta}}\cos(\beta_{12}z) + \frac{\sqrt{\Delta}+\delta}{\sqrt{\Delta}}\cos(\beta_{13}z) - \frac{\Delta-\delta^2}{\Delta}\cos(\beta_{23}z)\right] \end{cases} \quad (A1)$$

where $I_i(z)$ is the propagation distance dependent intensity in WG $i$, $i = 0, 1, 2$ and

$$\begin{cases} \Delta = 8\kappa^2 + \delta^2 \\ \beta_{ij} = \beta_i - \beta_j \\ \beta_1 = -\delta \\ \beta_2 = -\frac{\delta - \sqrt{\Delta}}{2} \\ \beta_3 = -\frac{\delta + \sqrt{\Delta}}{2} \end{cases}, \quad (A2)$$

where $\delta$ is the detuning, $\kappa$ is the coupling ratio, $\beta_i$ is the propagation constant for the $i^{th}$ supermode.

Assume the effective refractive indices for a-GST and c-GST are $n_a$ and $n_c$, respectively. The detuning between c- and a-GST, $\delta_{ac}$, is given by

$$\delta_{ca} = \frac{2\pi}{\lambda_0} \cdot (n_c - n_a) = k_0 \cdot dn_{ca}, \tag{A4}$$

where $dn_{ca} = n_c - n_a$. Assume $\delta_{ca} = \delta_c - \delta_a = (N-1)\delta$ and $\delta_a = \delta = M\kappa$, we can write down the following relations:

$$\begin{cases} \kappa = \frac{dn_{ca}}{M(N-1)} k_0 \\ \delta_a = M\kappa = \frac{dn_{ca}}{N-1} k_0 \\ \delta_c = N\delta_a = \frac{N \cdot dn_{ca}}{N-1} k_0 \\ L_0 = \frac{\pi}{2\sqrt{\Delta}} = \frac{M(N-1)}{4\sqrt{8+M^2} dn_{ca}} \lambda_0 \end{cases}, \tag{A5}$$

where $L_0$ is the coupling length of WG 0. The total device length is determined by $2L_0 \times$ (number of levels – 1) (as in Fig. 6). For example, if $M = 1.5$, the device length $L_{dev} = 2L_0 \times (3 - 1) = 4L_0$.

*A.2 Design procedures for PD-TWDs*

The design procedure can be described as the following steps:

1. Assume an operation wavelength of $\lambda_0 = 1.55 \, \mu m$ and standard 220 nm silicon on insulator wafer. The waveguide is formed by partially etching the silicon by 120 nm. The remaining 100 nm slab is used for doping. Determine suitable constants $M$ and $N$ to achieve reasonable numbers of levels, phase mismatch between a- and c-GST, and device length. Here we used $M = 1.5$ and $N = 30$ to achieve 3 operation levels and reasonable extinction ratio between a- and c-GST.

2. Fix the width of the WG 0 ($w_h$) and PCM thickness ($h_{pcm}$), such as to $w_h = 500$ nm and $h_{pcm} = 20 \, nm$. Then calculate the effective indices for both a- and c-phases to determine $dn_{ca}$. Calculate the desired detuning rate $\delta_a$, coupling coefficient $\kappa$ from $M$, $N$, $dn_{ca}$, and $\lambda_0$. Estimate the device length $L_{dev}$. If the device length is too long, increase PCM thickness ($h_{pcm}$) and repeat this step. Here, we used $h_{pcm} = 30\ nm$ to achieve a high refractive index contrast, and obtained the following parameters: $\delta_a = 0.04959\ \mu m^{-1}$, $\kappa = 0.03306\ \mu m^{-1}$ and $L_{dev} = 115\ \mu m$.

3. Sweep the width of the bare silicon WG ($w_b$) to optimize the detuning rate $\delta_a$. We calculated the desired effective index as 2.67177 and optimized $w_b = 630\ nm$.

4. Sweep the gap to optimize the coupling coefficient $\kappa$. We sweep the gap between 300 and 400 nm and obtained the desired $\kappa$ when it is 315 nm.

5. Run an FDTD simulation to verify the field distribution in the waveguide. Our FDTD results are shown in Fig. 8 and Fig. 9.

**Disclosures**

Authors declare that they have no competing interests.

**Code, Data, and Materials Availability**

The code for numerically solving the three-waveguide propagation problem can be found at https://github.com/charey6/TriWG.git. All data in support of the findings of this paper are available within the article or as appendix.

**Acknowledgments**

The research is funded by National Science Foundation (NSF-2003509), ONR-YIP Award, DRAPER Labs, DARPA-YFA Award, and Intel. Part of this work was conducted at the Washington Nanofabrication Facility/ Molecular Analysis Facility, a National Nanotechnology Coordinated Infrastructure (NNCI) site at the University of Washington with partial support from the National Science Foundation via awards NNCI-1542101 and NNCI-2025489.

**Author contributions**

R.C. and A.M. conceived the project. R.C. simulated the nonvolatile programmable units, fabricated the samples, and performed optical characterizations and data analysis. J.D., V.T., Z.F. and J.Z. helped with the design and data analysis. A.M. supervised and planned the project. R.C. wrote the manuscript with input from all the authors.

[28] Fang, Z., Mills, B., Chen, R., Zhang, J., Xu, P., Hu, J. and Majumdar, A., "Arbitrary Programming of Racetrack Resonators Using Low-Loss Phase-Change Material Sb2Se3," Nano Lett. (2023).
[29] Kim, S., Burr, G. W., Kim, W. and Nam, S.-W., "Phase-change memory cycling endurance," MRS Bulletin **44**(9), 710–714 (2019).
[30] Hudgens, S. and Johnson, B., "Overview of Phase-Change Chalcogenide Nonvolatile Memory Technology," MRS Bulletin **29**(11), 829–832 (2004).
[31] Zhang, Q., Zhang, Y., Li, J., Soref, R., Gu, T. and Hu, J., "Broadband nonvolatile photonic switching based on optical phase change materials: beyond the classical figure-of-merit," Optics Letters **43**(1), 94 (2018).
[32] Tuma, T., Pantazi, A., Le Gallo, M., Sebastian, A. and Eleftheriou, E., "Stochastic phase-change neurons," Nature Nanotechnology **11**(8), 693–699 (2016).
[33] Erickson, J. R., Shah, V., Wan, Q., Youngblood, N. and Xiong, F., "Designing fast and efficient electrically driven phase change photonics using foundry compatible waveguide-integrated microheaters," Optics Express **30**(8), 13673 (2022).
[34] Meng, J., Gui, Y., Nouri, B. M., Ma, X., Zhang, Y., Popescu, C.-C., Kang, M., Miscuglio, M., Peserico, N., Richardson, K., Hu, J., Dalir, H. and Sorger, V. J., "Electrical programmable multilevel nonvolatile photonic random-access memory," 1, Light Sci Appl **12**(1), 189 (2023).
[35] Zhang, Y., Chou, J. B., Li, J., Li, H., Du, Q., Yadav, A., Zhou, S., Shalaginov, M. Y., Fang, Z., Zhong, H., Roberts, C., Robinson, P., Bohlin, B., Ríos, C., Lin, H., Kang, M., Gu, T., Warner, J., Liberman, V., et al., "Broadband transparent optical phase change materials for high-performance nonvolatile photonics," Nature Communications **10**(1), 4279 (2019).
[36] Yang, X., Nisar, M. S., Nisar, M. S., Yuan, W., Zheng, F., Lu, L., Lu, L., Chen, J., Chen, J., Zhou, L. and Zhou, L., "Phase change material enabled 2 × 2 silicon nonvolatile optical switch," Opt. Lett., OL **46**(17), 4224–4227 (2021).
[37] Chen, Y., Ho, S.-T. and Krishnamurthy, V., "All-optical switching in a symmetric three-waveguide coupler with phase-mismatched absorptive central waveguide," Applied Optics **52**(36), 8845 (2013).
[38] Teo, T. Y., Teo, T. Y., Krbal, M., Mistrik, J., Mistrik, J., Prikryl, J., Lu, L., Simpson, R. E. and Simpson, R. E., "Comparison and analysis of phase change materials-based reconfigurable silicon photonic directional couplers," Opt. Mater. Express, OME **12**(2), 606–621 (2022).
**Rui Chen** is a 4th year graduate student at the University of Washington working with Prof. Arka Majumdar. He received his bachelor's degree from Zhejiang University (2018) and his M.S. degree (2020) from Columbia University. His main research interests include phase-change materials based reconfigurable photonic integrated circuits and programmable metasurfaces.

**Arka Majumdar** is an associate professor at the Departments of Electrical and Computer Engineering and Physics, University of Washington. He received B. Tech. degree from IIT-Kharagpur (2007), where he was honored with the President's Gold Medal. He completed his M.S. degree (2009) and Ph.D. degree (2012) in Electrical Engineering at Stanford University. He spent one year at the University of California, Berkeley (2012–2013) as a postdoc before joining Intel Labs (2013–2014). His research interest includes developing a hybrid nanophotonic platform using emerging material systems for optical information science, imaging, and microscopy.

Biographies and photographs for the other authors are not available.